\documentclass[a4paper]{jpconf}
\usepackage{graphicx}
\newcommand{\beq}{\begin{equation}}
\newcommand{\eeq}{\end{equation}}
\newcommand{\bea}{\begin{eqnarray}}
\newcommand{\eea}{\end{eqnarray}}
\newcommand \hmu {\hat{\mu}}
\begin{document}
\title{Strangeness at finite temperature from Lattice QCD}

\author{J. Noronha-Hostler$^a$, R. Bellwied$^a$, J. G\"unther$^b$, P. Parotto$^a$, A. Pasztor$^b$, I. Portillo Vazquez$^a$, C. Ratti$^a$}

\address{\small{\it $^a$ Department of Physics, University of Houston, Houston, TX 77204, USA}}
\address{\small{\it $^b$ Department of Physics, University of Wuppertal, Gaussstr. 20, D-42119 Wuppertal, Germany}}

\ead{jakinoronhahostler@gmail.com}

\begin{abstract}
The precision reached by recent lattice QCD results allows for the first time to investigate whether the measured hadronic spectrum is missing some additional strange states, which are predicted by the Quark Model  but have not yet been detected. This can be done by comparing some sensitive thermodynamic observables from lattice QCD to the predictions of the Hadron Resonance Gas model (with the inclusion of decays [3]). We propose a set of specific observables, defined as linear combinations of conserved charge fluctuations, which allow to investigate this issue for baryons containing one or more strange quarks separately. Applications of these observables to isolate the multiplicity fluctuations of kaons from lattice QCD, and their comparison with the experimental results, are also discussed.
\end{abstract}

\section{Introduction}
In the 1960's Ralf Hagedorn proposed  \cite{Hagedorn:1965st} that if there was a limiting temperature of the universe, now known as the Hagedorn Temperature, then the addition of increasingly more energy to a system would no longer increase the temperature but rather  create more massive, highly degenerate resonances.  The consequence of this idea was an exponentially increasing mass spectrum 
\begin{equation}\label{eqn:mass}
N(M)=\sum_i d_i\Theta_i(M-M_i)
\end{equation}
summed over the degeneracy, $d_i$, of the known hadrons.  In 2004 \cite{Broniowski:2004yh} and 2015 \cite{Lo:2015cca} the  experimentally measured hadrons from the Particle Data Group \cite{Agashe:2014kda} confirmed the continually exponentially increasing mass spectrum, as Hagedorn originally suggested. 

Meanwhile,  high energy heavy-ion collisions at RHIC and the LHC probed temperatures surpassing Hagedorn's original limiting temperature, producing a deconfined state of matter known as the Quark Gluon Plasma. We now understand, thanks to first principle Lattice QCD calculations, that there is a cross-over phase transition \cite{Aoki:2006we}, not a limiting temperature.  In this framework we can understand the Hagedorn temperature as roughly equivalent to the critical temperature and then expect the effect of an exponentially increasing mass spectrum to appear close to the phase transition.  Indeed, including missing resonances close to the phase transition can affect dynamical chemical equilibrium \cite{NoronhaHostler:2007jf,NoronhaHostler:2009cf,Beitel:2014kza,Beitel:2016ghw}, decrease the shear viscosity over entropy ratio  \cite{NoronhaHostler:2008ju,NoronhaHostler:2012ug,Pal:2010es,Kadam:2014cua}, affect the elliptical flow \cite{Noronha-Hostler:2013rcw,Paquet:2015lta}, and improve thermal fits \cite{NoronhaHostler:2009tz}. 

Recent comparisons to Lattice Quantum Chromodynamic calculations \cite{Bazavov:2014xya} suggested that there may be missing strange hadrons as calculated from Quark Model states  \cite{Capstick:1986bm,Ebert:2009ub} due to a mismatch in the strange chemical potential to baryon chemical potential in Lattice QCD vs. the Hadron Resonance gas model from the known PDG spectrum \cite{Agashe:2014kda}. Further more, there were suggestions \cite{Bazavov:2014xya,Noronha-Hostler:2014usa,Noronha-Hostler:2014aia} that missing resonances could account for the $p/\pi$ vs. strange hadron tension at LHC when it comes to the thermal fits \cite{Abelev:2012wca}.  However, in \cite{Bazavov:2014xya} the decays of the Quark Model states were not considered, which are necessary for thermal fits. 

\begin{figure}[h]
\begin{minipage}{18pc}
\includegraphics[width=18pc]{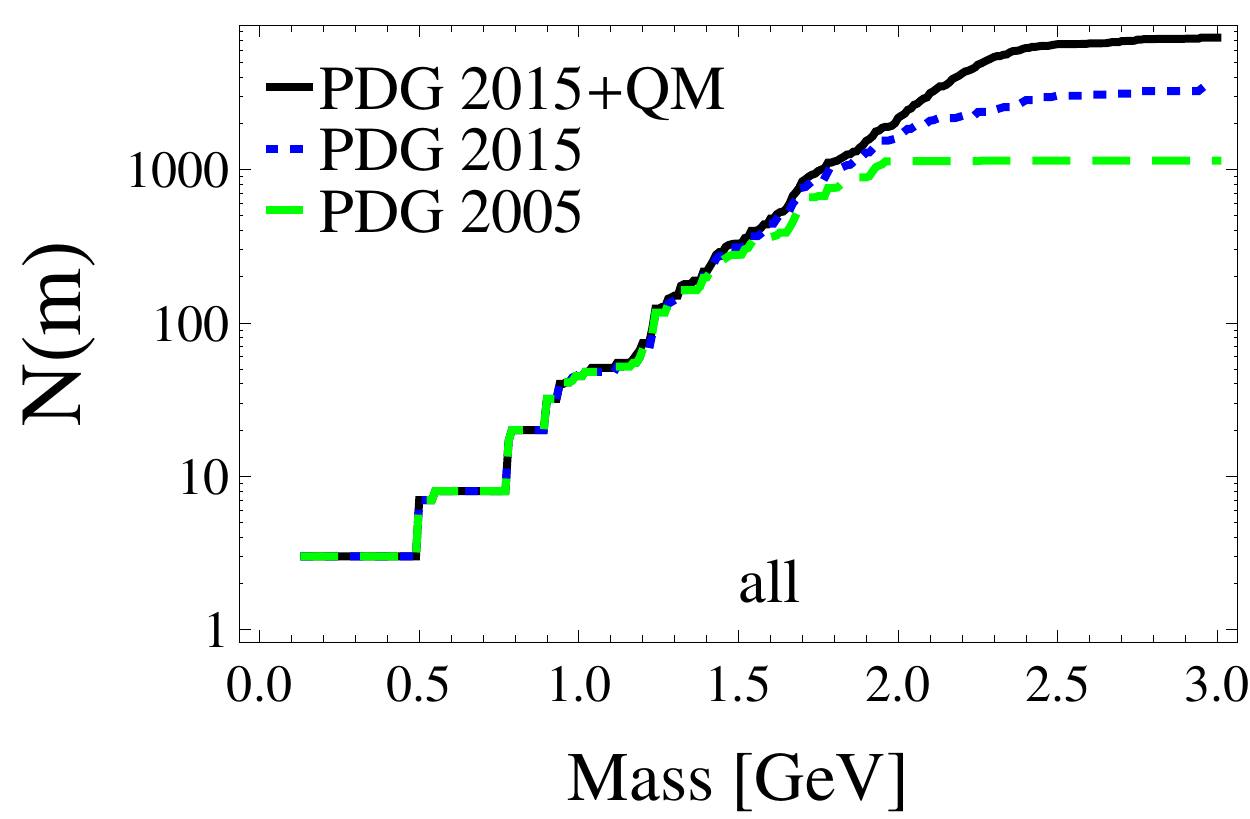}
\caption{\label{all}Mass spectrum, Eq. (\ref{eqn:mass}), of the strange mesons for the PDG05, PDG15, and PDG15+Quark Model states. }
\end{minipage}\hspace{2pc}%
\begin{minipage}{18pc}
\includegraphics[width=18pc]{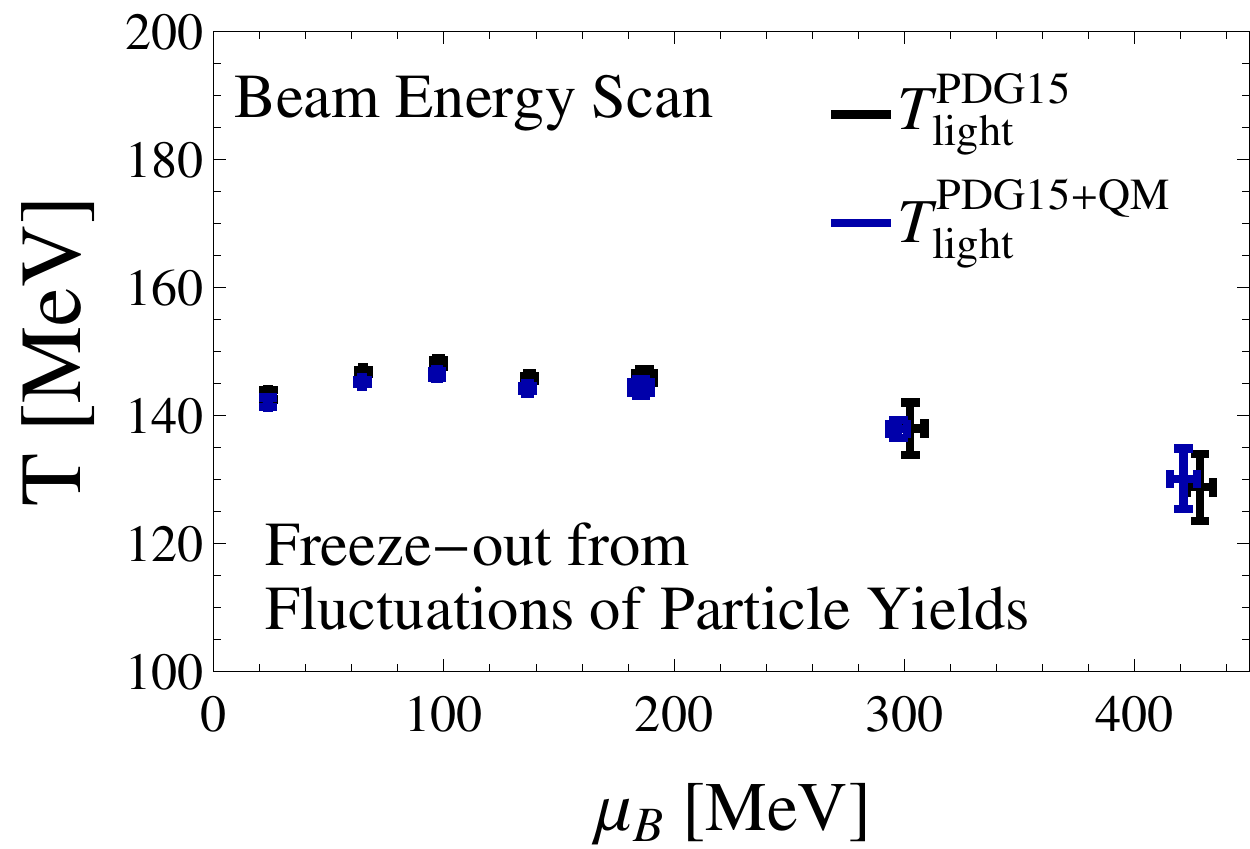}
\caption{\label{sup}Comparison of charged strange mesons from Lattice QCD via the partial pressure to the experimentally measured (preliminary) kaon fluctuations at STAR for $\sqrt{s_{NN}}=200$ GeV.}
\end{minipage} 
\end{figure}

Fig. \ref{all} shows the exponentially increasing mass spectra that includes these Quark Model states, implying that these missing resonances are consistent with Hagedorn's original postulate.  Using the known branching ratios from \cite{Agashe:2014kda}, we extrapolated up the branching ratios of the Quark Model states taking all quantum numbers into account.  Following this, we analyzed the net-proton and net-charge fluctuations ($\chi_1/\chi_2$) as in \cite{Alba:2014eba}, to extract the corresponding $T$ and $\mu_B$ across energies in the Beam Energy Scan.  The results are shown in Fig. \ref{sup} where we find that the addition of the Quark Model states slightly decreases the freeze-out temperature but overall only has a small affect.  This is consistent with previous studies analyzing the affect of additional resonances on the thermal fits \cite{NoronhaHostler:2009tz}. 

An alternative picture was also suggested to resolve the tension between the light and strange hadrons at LHC. In Lattice QCD, the inflection point of susceptibilities can provide clues of about the temperature of hadronization.  Using the light susceptibility one finds an inflection point around $T\sim 150$ MeV whereas the inflection point for the strangeness susceptibility is around $T\sim 165$ MeV \cite{Bellwied:2013cta}.  Considering there is $\sim15$ MeV difference between light and strange hadrons, a logical consequence of this may be that strange hadrons reach chemical equilibrium at higher temperatures than light hadrons.  If this is true, it would be consistent with the tension between the light and strange hadrons because it would increase the population of strange baryons, which are typically under-predicted using lower temperatures. 

This idea is consistent with many dynamical models. In UrQMD there is no specified chemical freeze-out temperature such that each particle species reaches chemical equilibrium on a different time scale \cite{Bass:1999tu}.  Similarly, using multi-body hadronic interactions via rate equations, one can also reach chemical equilibrium on different time scales depending on the species.  However, if one can provide directly from first-principle Lattice calculations that different chemical equilibration temperatures are needed then it gives significantly more weight to these dynamical models.  Furthermore, it will then require the hadronization schemes to be updated uniformly to include different temperatures for light and strange hadrons.  

In order to extract the chemical freeze-out temperature from the lattice we use ratios of susceptibilities as discussed in  \cite{Alba:2014eba,Karsch:2012wm,Borsanyi:2014ewa}.  In the experiment, the only strange multiplicity fluctuations (and their corresponding moments of these distributions) that can be currently measured at the Beam Energy Scan are charged kaons \cite{JXU}.  However, on the lattice all particles exist as well as their corresponding interactions. In order to extract only the charged kaon contribution, we implement partial pressures for charged strange mesons as in \cite{Noronha-Hostler:2016rpd}
$\frac{\chi^K_2}{\chi^K_1}=\frac{\cosh(\hmu_S+\hmu_Q)}{\sinh(\hmu_S+\hmu_Q)}$
where $\hmu_S$ and $\hmu_Q$ are supplied from the Wuppertal Budapest collaboration.

\begin{figure}[h]
\centering
\includegraphics[width=20pc]{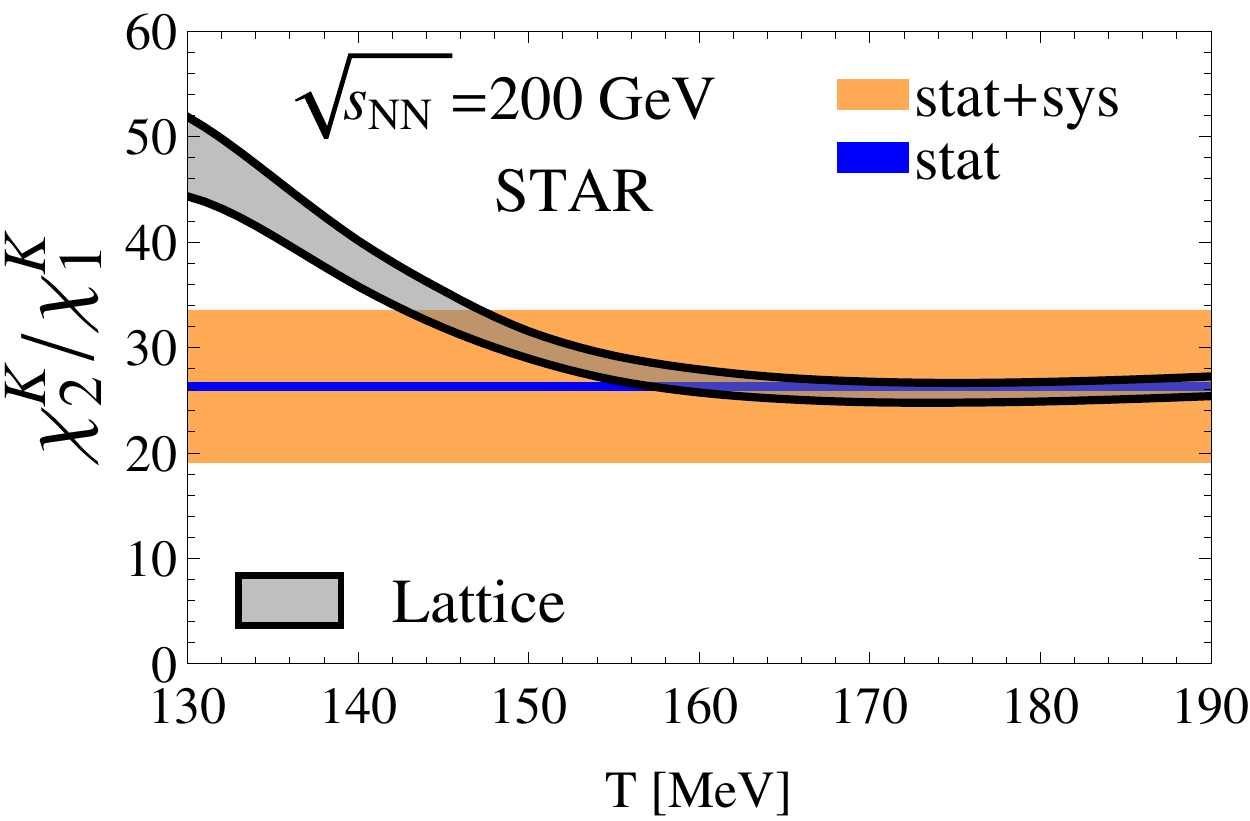}
\caption{\label{kaons}Comparison of net-$K^{+/-}$ from Lattice QCD (Wuppertal-Budapest)  to preliminary STAR  net-$K^{+/-}$ for $\sqrt{s_{NN}}=200$ GeV.}
\end{figure}

In Fig\ \ref{kaons} a comparison between the Lattice QCD kaon partial pressure  is shown compared to the preliminary STAR data for $\frac{\langle |K^{+/-}|^2\rangle}{\langle K^{+/-}\rangle}$.  Due to the large error bars on the kaons data, it is not yet possible to extract the strange chemical freeze-out temperature from Lattice QCD.  However, if the error bars are decreased in the future then we will be able to say decisively if there is a splitting between the light and strange chemical equilibration temperatures.

In this proceedings, we introduce two approaches to study the possible differences in chemical equilibration temperatures between light and strange hadrons. The first approach  is to include additional states predicted from the Quark Model as well as modeling their decay channels and branching ratios to see if this fixes the tension between light and strange hadrons. The second approach is a new method to extract the charged, strange susceptibilities from Lattice QCD and extract the strange freeze-out temperature from first principles.  At this point in time, the error bars on kaon fluctuations are too large to extract the strange freeze-out temperature.  However, if the error bars are decreased in the future we may be able to decidedly settle the tension between light and strange hadrons.

\section{Acknowledgements}
This material is based upon work supported by the National Science Foundation under grant no. PHY-1513864 and by the U.S. Department of Energy, Office of Science, Office of Nuclear Physics, within the framework of the Beam Energy Scan Theory (BEST) Topical Collaboration.  This work contains lattice QCD data provided by the Wuppertal-Budapest Collaboration. An  award  of  computer  time  was  provided
by the INCITE program.  This research used resources of the Argonne Leadership Computing Facility, which is a DOE Office of Science User Facility supported under Contract DE-AC02-06CH11357. The work of J. G. and A. P. was supported by the DFG grant SFB/TR55. The authors gratefully acknowledge the Gauss Centre for Supercomputing (GCS) for providing computing time for a GCS Large-Scale Project on the GCS share of the supercomputer JUQUEEN \cite{juqueen} at J\"ulich Supercomputing Centre (JSC).


\section*{References}
\begin{thebibliography}{9}

\bibitem{Hagedorn:1965st} 
  R.~Hagedorn,
  Nuovo Cim.\ Suppl.\  {\bf 3}, 147 (1965).

\bibitem{Broniowski:2004yh} 
  W.~Broniowski, W.~Florkowski and L.~Y.~Glozman,
  Phys.\ Rev.\ D {\bf 70}, 117503 (2004)
  doi:10.1103/PhysRevD.70.117503
  [hep-ph/0407290].
  
\bibitem{Lo:2015cca} 
  P.~M.~Lo, M.~Marczenko, K.~Redlich and C.~Sasaki,
  Phys.\ Rev.\ C {\bf 92}, no. 5, 055206 (2015)
  doi:10.1103/PhysRevC.92.055206
  [arXiv:1507.06398 [nucl-th]].
  
\bibitem{Agashe:2014kda} 
  K.~A.~Olive {\it et al.} [Particle Data Group Collaboration],
  Chin.\ Phys.\ C {\bf 38}, 090001 (2014).
  doi:10.1088/1674-1137/38/9/090001
  
\bibitem{Aoki:2006we} 
  Y.~Aoki, G.~Endrodi, Z.~Fodor, S.~D.~Katz and K.~K.~Szabo,
  Nature {\bf 443}, 675 (2006)
  doi:10.1038/nature05120
  [hep-lat/0611014].



\bibitem{NoronhaHostler:2007jf} 
  J.~Noronha-Hostler, C.~Greiner and I.~A.~Shovkovy,
  Phys.\ Rev.\ Lett.\  {\bf 100}, 252301 (2008)
  doi:10.1103/PhysRevLett.100.252301
  [arXiv:0711.0930 [nucl-th]].
\bibitem{NoronhaHostler:2009cf} 
  J.~Noronha-Hostler, M.~Beitel, C.~Greiner and I.~Shovkovy,
  Phys.\ Rev.\ C {\bf 81}, 054909 (2010)
  doi:10.1103/PhysRevC.81.054909
  [arXiv:0909.2908 [nucl-th]].
\bibitem{Beitel:2014kza} 
  M.~Beitel, K.~Gallmeister and C.~Greiner,
  Phys.\ Rev.\ C {\bf 90}, no. 4, 045203 (2014)
  doi:10.1103/PhysRevC.90.045203
  [arXiv:1402.1458 [hep-ph]].
\bibitem{Beitel:2016ghw} 
  M.~Beitel, C.~Greiner and H.~Stoecker,
  Phys.\ Rev.\ C {\bf 94}, no. 2, 021902 (2016)
  doi:10.1103/PhysRevC.94.021902
  [arXiv:1601.02474 [hep-ph]].
  

  
\bibitem{NoronhaHostler:2008ju} 
  J.~Noronha-Hostler, J.~Noronha and C.~Greiner,
  Phys.\ Rev.\ Lett.\  {\bf 103}, 172302 (2009)
  doi:10.1103/PhysRevLett.103.172302
  [arXiv:0811.1571 [nucl-th]].
\bibitem{NoronhaHostler:2012ug} 
  J.~Noronha-Hostler, J.~Noronha and C.~Greiner,
  Phys.\ Rev.\ C {\bf 86}, 024913 (2012)
  doi:10.1103/PhysRevC.86.024913
  [arXiv:1206.5138 [nucl-th]].
\bibitem{Kadam:2014cua} 
  G.~P.~Kadam and H.~Mishra,
  Nucl.\ Phys.\ A {\bf 934}, 133 (2014)
  doi:10.1016/j.nuclphysa.2014.12.004
  [arXiv:1408.6329 [hep-ph]].
\bibitem{Pal:2010es} 
  S.~Pal,
  Phys.\ Lett.\ B {\bf 684}, 211 (2010)
  doi:10.1016/j.physletb.2010.01.017
  [arXiv:1001.1585 [nucl-th]].
  

  
  
\bibitem{Noronha-Hostler:2013rcw} 
  J.~Noronha-Hostler, J.~Noronha, G.~S.~Denicol, R.~P.~G.~Andrade, F.~Grassi and C.~Greiner,
  Phys.\ Rev.\ C {\bf 89}, no. 5, 054904 (2014)
  doi:10.1103/PhysRevC.89.054904
  [arXiv:1302.7038 [nucl-th]].
\bibitem{Paquet:2015lta} 
  J.~F.~Paquet, C.~Shen, G.~S.~Denicol, M.~Luzum, B.~Schenke, S.~Jeon and C.~Gale,
  Phys.\ Rev.\ C {\bf 93}, no. 4, 044906 (2016)
  doi:10.1103/PhysRevC.93.044906
  [arXiv:1509.06738 [hep-ph]].
  
  

  
  
\bibitem{NoronhaHostler:2009tz} 
  J.~Noronha-Hostler, H.~Ahmad, J.~Noronha and C.~Greiner,
  Phys.\ Rev.\ C {\bf 82}, 024913 (2010)
  doi:10.1103/PhysRevC.82.024913
  [arXiv:0906.3960 [nucl-th]].
  
 
  
\bibitem{Capstick:1986bm} 
  S.~Capstick and N.~Isgur,
  Phys.\ Rev.\ D {\bf 34}, 2809 (1986)
  [AIP Conf.\ Proc.\  {\bf 132}, 267 (1985)].
  doi:10.1103/PhysRevD.34.2809, 10.1063/1.35361
  
\bibitem{Ebert:2009ub} 
  D.~Ebert, R.~N.~Faustov and V.~O.~Galkin,
  Phys.\ Rev.\ D {\bf 79}, 114029 (2009)
  doi:10.1103/PhysRevD.79.114029
  [arXiv:0903.5183 [hep-ph]].

\bibitem{Bazavov:2014xya} 
  A.~Bazavov {\it et al.},
  Phys.\ Rev.\ Lett.\  {\bf 113}, no. 7, 072001 (2014)
  doi:10.1103/PhysRevLett.113.072001
  [arXiv:1404.6511 [hep-lat]].
  
\bibitem{Noronha-Hostler:2014usa} 
  J.~Noronha-Hostler and C.~Greiner,
  arXiv:1405.7298 [nucl-th].
\bibitem{Noronha-Hostler:2014aia} 
  J.~Noronha-Hostler and C.~Greiner,
  Nucl.\ Phys.\ A {\bf 931}, 1108 (2014)
  doi:10.1016/j.nuclphysa.2014.08.101
  [arXiv:1408.0761 [nucl-th]].


\bibitem{Abelev:2012wca} 
  B.~Abelev {\it et al.} [ALICE Collaboration],
  Phys.\ Rev.\ Lett.\  {\bf 109}, 252301 (2012)
  doi:10.1103/PhysRevLett.109.252301
  [arXiv:1208.1974 [hep-ex]].
  
  
\bibitem{Alba:2014eba} 
  P.~Alba {\it et al.} ,
  Phys.\ Lett.\ B {\bf 738}, 305 (2014)
  doi:10.1016/j.physletb.2014.09.052
  [arXiv:1403.4903 [hep-ph]].
  


\bibitem{Bellwied:2013cta} 
  R.~Bellwied, S.~Borsanyi, Z.~Fodor, S.~D.~Katz and C.~Ratti,
  Phys.\ Rev.\ Lett.\  {\bf 111}, 202302 (2013)
  doi:10.1103/PhysRevLett.111.202302
  [arXiv:1305.6297 [hep-lat]].
  
\bibitem{Bass:1999tu} 
  S.~A.~Bass {\it et al.} ,
  Phys.\ Rev.\ C {\bf 60}, 021902 (1999)
  doi:10.1103/PhysRevC.60.021902
  [nucl-th/9902062].
  
\bibitem{Karsch:2012wm} 
  F.~Karsch,
  Central Eur.\ J.\ Phys.\  {\bf 10}, 1234 (2012)
  doi:10.2478/s11534-012-0074-3
  [arXiv:1202.4173 [hep-lat]].
  
\bibitem{Borsanyi:2014ewa} 
  S.~Borsanyi, Z.~Fodor, S.~D.~Katz, S.~Krieg, C.~Ratti and K.~K.~Szabo,
  Phys.\ Rev.\ Lett.\  {\bf 113}, 052301 (2014)
  doi:10.1103/PhysRevLett.113.052301
  [arXiv:1403.4576 [hep-lat]].
  
  \bibitem{JXU} 
  J.~Xu for [STAR], Strangeness in Quark Matter 2016 proceedings

\bibitem{Noronha-Hostler:2016rpd} 
  J.~Noronha-Hostler, R.~Bellwied, J.~Gunther, P.~Parotto, A.~Pasztor, I.~P.~Vazquez and C.~Ratti,
  arXiv:1607.02527 [hep-ph].
  

  
\bibitem{juqueen}
  JUQUEEN: IBM Blue Gene/Q Supercomputer System at the J{\"u}lich Supercomputing Centre
  JUQUEEN: IBM Blue Gene/Q Supercomputer
System at the J{\"u}lich Supercomputing Centre,
Tech. Rep. 1 A1 (J{\"u}lich Supercomputing Centre,
http://dx.doi.org/10.17815/jlsrf-1-18, 2015).
  

  
\end{thebibliography}
\end{document}